\documentclass[12pt]{article}
\usepackage{epsfig}
\date{}
\begin{document}

{\large \sf
\title{
{\normalsize
\begin{flushright}
\end{flushright}}
{\vspace{2cm} {\LARGE \sf Theory of Timeon\thanks{This research was
supported in part by the U.S. Department of Energy (grant no.
DE-FG02-92-ER40699)}} \vspace{2cm}} }

{\large \sf
\author{
{\large \sf
R. Friedberg$^1$ and  T. D. Lee$^{1,~2}$}\\
{\normalsize \it 1. Physics Department, Columbia University}\\
{\normalsize \it New York, NY 10027, U.S.A.}\\
{\normalsize \it 2. China Center of Advanced Science and Technology (CCAST/World Lab.)}\\
{\normalsize \it P.O. Box 8730, Beijing 100080, China}\\
} \maketitle

\begin{abstract}

{\normalsize \sf

It is proposed that $T$ violation in hadronic physics, as well as
the masses of $u,~d$ quarks, arise from a pseudoscalar interaction
with a new spin $0$ field $\tau(x)$, odd in $P$ and $T$, but even in
$C$. This interaction contains a factor $i\gamma_5$ in the quark
Dirac algebra, so that the full Hamiltonian is $P$, $T$ conserving;
but by spontaneous symmetry breaking, the new field $\tau(x)$ has a
nonzero expectation value $<\tau>=\tau_0$ that breaks $P$ and $T$
symmetry.

Oscillations of $\tau(x)$ about its expectation value $\tau_0$
produce a new particle, the "timeon", whose mass is  independent of
any known quantities. If the timeon mass is within the range of
present accelerators, observation of the particle can be helped with
a search of $T$-violating events. }
\end{abstract}

{\normalsize \sf PACS{:~~12.15.Ff,~~11.30.Er}}

\vspace{1cm}


\newpage

\section*{\Large \sf  1. Introduction}
\setcounter{section}{1} \setcounter{equation}{0}

We assume that the standard model of weak interaction[1] is $CP$ and
$T$ conserving. In this paper we propose that the observed $CP$ and
$T$ violations are due to a new $P$ odd and $T$ odd spin zero field
$\tau(x)$, called the timeon field; the same field is also
responsible for the small masses of $u,~d$ quarks, as well as for
those of $\nu_1$ and the electron. In the following, our discussions
will be restricted only to quarks.

Let $q_i(\uparrow)$ and $q_i(\downarrow)$ be the quark states
"diagonal" in $W^\pm$ transitions[2]:
$$
q_i(\downarrow)\rightleftharpoons q_i(\uparrow) + W^-\eqno(1.1)
$$
and
$$
q_i(\uparrow)\rightleftharpoons q_i(\downarrow) + W^+\eqno(1.2)
$$
with $i=1,~2$ and $3$. The electric charges in units of $e$ are
$+\frac{2}{3}$ for $q_i(\uparrow)$ and $-\frac{1}{3}$ for
$q_i(\downarrow)$. These quark states $q_i(\uparrow)$ and
$q_i(\downarrow)$ are, however, not the mass eigenstates $d,~s,~b$
and $u,~c,~t$. We assume that the mass Hamiltonians $H_\uparrow$ for
$q_i(\uparrow)$ and $H_\downarrow$ for $q_i(\downarrow)$ are given
by
$$
H_{\uparrow/\downarrow}=\bigg(q_1^\dag,~
q_2^\dag,~q_3^\dag\bigg)_{\uparrow/\downarrow} (G \gamma_4 +
iF\gamma_4 \gamma_5)_{\uparrow/\downarrow}\left(
\begin{array}{r}
q_1\\
q_2\\
q_3
\end{array}\right)_{\uparrow/\downarrow}\eqno(1.3)
$$
with the $3\times 3$ matrices $G_{\uparrow/\downarrow}$ and
$F_{\uparrow/\downarrow}$ both real and hermitian. The mass matrix
$G_{\uparrow/\downarrow}$ is the same zeroth order mass matrix as
${\cal M}_0(q_{\uparrow/\downarrow})$ of Ref.~2, given by
$$
G_{\uparrow/\downarrow}=\left(
\begin{array}{ccc}
\beta \eta^2(1+\xi^2)& -\beta \eta & -\beta \xi \eta\\
-\beta \eta &\beta + \alpha \xi^2 & -\alpha \xi\\
-\beta \xi \eta &-\alpha \xi& \alpha +\beta
\end{array}\right)_{\uparrow/\downarrow}, \eqno(1.4)
$$
with $\alpha_\uparrow,~\beta_\uparrow,~ \xi_\uparrow,~\eta_\uparrow$
and $\alpha_\downarrow,~\beta_\downarrow,~
\xi_\downarrow,~\eta_\downarrow$ all real parameters. It can be
readily verified that the determinants
$$
|G_\uparrow|=|G_\downarrow|=0.\eqno(1.5)
$$
We assume $\alpha_{\uparrow/\downarrow}$ and
$\beta_{\uparrow/\downarrow}$ to be all positive. The lowest
eigenvalues of $G_\uparrow$ and $G_\downarrow$ are then both zero.
These two real symmetric matrices can be readily diagonalized by
real, orthogonal matrices $(U_\uparrow)_0$ and $(U_\downarrow)_0$,
with
$$
(U_\uparrow)_0^\dag G_\uparrow (U_\uparrow)_0=\left(
\begin{array}{ccc}
0 & 0 & 0\\
0 & m_0(c) & 0\\
0 & 0 & m_0(t)
\end{array}\right) \eqno(1.6)
$$
and
$$
(U_\downarrow)_0^\dag G_\downarrow (U_\downarrow)_0=\left(
\begin{array}{ccc}
0 & 0 & 0\\
0 & m_0(s) & 0\\
0 & 0 & m_0(b)
\end{array}\right) \eqno(1.7)
$$
where the nonzero eigenvalues are the zeroth order masses of $c,~t$
and $s,~b$ quarks, with
$$
m_0(c)=\beta_\uparrow[1+\eta_\uparrow^2(1+\xi_\uparrow^2)],\eqno(1.8)
$$
$$
m_0(t)= \alpha_\uparrow(1+\xi_\uparrow^2)+\beta_\uparrow,\eqno(1.9)
$$
$$
m_0(s)=\beta_\downarrow[1+\eta_\downarrow^2(1+\xi_\downarrow^2)]\eqno(1.10)
$$
and
$$
 m_0(b)=\alpha_\downarrow(1+\xi_\downarrow^2)+\beta_\downarrow.\eqno(1.11)
$$
Thus, $G_\uparrow$ and $G_\downarrow$ can be each represented by an
ellipse of minor and major axes given by $m_0(c)$ and $m_0(t)$ for
$\uparrow$ and likewise $m_0(s)$ and $m_0(b)$ for $\downarrow$.

The orientations of these two elliptic plates are determined by
their eigenstates. As in Ref.~2, we define four real angular
variables $\theta_\downarrow,~\phi_\downarrow$ and
$\theta_\uparrow,~\phi_\uparrow$ by
$$
\xi_\downarrow=\tan \phi_\downarrow,~~\xi_\uparrow=\tan
\phi_\uparrow
$$
$$
\eta_\downarrow=\tan \theta_\downarrow \cos \phi_\downarrow~~{\sf
and}~~\eta_\uparrow=\tan \theta_\uparrow \cos \phi_\uparrow.
\eqno(1.12)
$$
The eigenstates of $G_\uparrow$ are
$$
\epsilon_\uparrow = \left(
\begin{array}{l}
\cos \theta_\uparrow\\
\sin \theta_\uparrow \cos \phi_\uparrow\\
\sin \theta_\uparrow \sin \phi_\uparrow
\end{array}
\right)~{\sf with~eigenvalue}~0, \eqno(1.13)
$$
$$
p_\uparrow = \left(
\begin{array}{l}
-\sin \theta_\uparrow\\
\cos \theta_\uparrow \cos \phi_\uparrow\\
\cos \theta_\uparrow \sin \phi_\uparrow
\end{array}
\right)~{\sf with~eigenvalue}~m_0(c) \eqno(1.14)
$$
and
$$
P_\uparrow = \left(
\begin{array}{l}
~~~~0\\
-\sin \phi_\uparrow\\
~~\cos \phi_\uparrow
\end{array}
\right)~{\sf with~eigenvalue}~m_0(t). \eqno(1.15)
$$
Correspondingly, the eigenstates of $G_\downarrow$ are
$$
\epsilon_\downarrow = \left(
\begin{array}{l}
~~\cos \theta_\downarrow\\
-\sin \theta_\downarrow \cos \phi_\downarrow\\
-\sin \theta_\downarrow \sin \phi_\downarrow
\end{array}
\right)~{\sf with~eigenvalue}~0, \eqno(1.16)
$$
$$
p_\downarrow = \left(
\begin{array}{l}
\sin \theta_\downarrow\\
\cos \theta_\downarrow \cos \phi_\downarrow\\
\cos \theta_\downarrow \sin \phi_\downarrow
\end{array}
\right)~{\sf with~eigenvalue}~m_0(s), \eqno(1.17)
$$
and
$$
P_\downarrow = \left(
\begin{array}{l}
~~~~0\\
-\sin \phi_\downarrow\\
~~\cos \phi_\downarrow
\end{array}
\right)~{\sf with~eigenvalue}~m_0(b). \eqno(1.18)
$$
We note that by changing $\theta_\uparrow$, $\phi_\uparrow$ to
$-\theta_\downarrow$, $\phi_\downarrow$ the unit vectors
$\epsilon_\uparrow$, $p_\uparrow$, $P_\uparrow$ of (1.13)-(1.15)
become $\epsilon_\downarrow$, $p_\downarrow$ and $P_\downarrow$ of
(1.16)-(1.18). Here the signs of $\theta_\uparrow$ and
$\theta_\downarrow$ are chosen so that the sign convention of the
particle data group's CKM matrix agrees with both $\theta_\uparrow$
and $\theta_\downarrow$ being positive, as we shall see. In terms of
these eigenstates, the $3\times 3$ real unitary matrices
$(U_\uparrow)_0$ and $(U_\downarrow)_0$ of (1.6)-(1.7) are given by
$$
(U_\downarrow)_0=(\epsilon_\downarrow,~p_\downarrow,~P_\downarrow)\eqno(1.19)
$$
and
$$
(U_\uparrow)_0=(\epsilon_\uparrow,~p_\uparrow,~P_\uparrow).\eqno(1.20)
$$
Thus, in the absence of the $iF_{\uparrow/\downarrow}
\gamma_4\gamma_5$ term in (1.3), the corresponding CKM matrix in
this approximation is given by
$$
(U_{CKM})_0=(U_\downarrow)_0^\dag (U_\uparrow)_0=
$$
$$\left(
\begin{array}{ccc}
\cos \theta_\downarrow\cos \theta_\uparrow &\sin\theta_\downarrow \cos\theta_\uparrow &\sin\theta_\uparrow \sin\phi\\
~~-\sin\theta_\downarrow \sin\theta_\uparrow \cos\phi
&~~+\cos\theta_\downarrow \sin\theta_\uparrow \cos\phi &\\
&&\\
-\cos \theta_\downarrow\sin \theta_\uparrow &-\sin\theta_\downarrow \sin\theta_\uparrow &\cos\theta_\uparrow \sin\phi\\
~~-\sin\theta_\downarrow \cos\theta_\uparrow \cos\phi
&~~+\cos\theta_\downarrow \cos\theta_\uparrow \cos\phi &\\
&&\\
 \sin\theta_\downarrow \sin\phi
&-\cos\theta_\downarrow\sin\phi&\cos\phi
\end{array}
 \right )~,\eqno(1.21)
$$
in which
$$
\phi=\phi_\uparrow-\phi_\downarrow.\eqno(1.22)
$$

The new hypothesis of this paper is to assume that $T$ violation and
the small masses of $u,~d$ quarks are due to the new
$$
iF\gamma_4\gamma_5\eqno(1.23)
$$
term in (1.3), with
$$
F_\uparrow=F_\downarrow=F=\tau_0f\tilde{f}\eqno(1.24)
$$
in which $\tau_0$ is a constant and $f$ a $3$ dimensional unit
vector represented by its $3\times 1$ real column matrix.
Graphically, we can visualize $G_\uparrow$ and $G_\downarrow$ as two
elliptic plates mentioned above, and $F$ as a single needle of
length $\tau_0$ and direction $f$, as shown in Figure~1.

As we shall discuss, the $F$-term (1.24) is due to the spontaneous
symmetry breaking of a new $T$ odd, $P$ odd and $CP$ odd, spin $0$
field $\tau(x)$, which has a vacuum expectation value given by
$$
<\tau(x)>_{vac}=\tau_0\neq 0.\eqno(1.25)
$$
While the general characteristics of spontaneous time reversal
symmetry breaking models have  been discussed in the literature[3],
the special new feature of the present model is to connect such
symmetry breaking with the smallness of up, down quark masses.

In Section 2, we begin with a general $T$, $P$ and $CP$ violating
mass matrix of the form
$$
G \gamma_4 + iF\gamma_4 \gamma_5\eqno(1.26)
$$
given by (1.3) with $G$ and $F$ both real and hermitian matrices,
then compare it with an alternative form of a single
$$
{\cal M}\gamma_4\eqno(1.27)
$$
term, but with ${\cal M}$ complex and hermitian. As will be shown,
these two different forms of mass matrices are in fact equivalent.
Differences appear when one goes beyond the mass matrices. In
Sections 3 and 4, we summarize the analysis of how in (1.24), the
length $\tau_0$ and the direction $f$ of the needle are related to
the light quark masses and the Jarlskog invariant[4] ${\cal J}$. As
we shall see, this leads to
$$
\tau_0\cong 33 MeV,\eqno(1.28)
$$
$$
m_u\cong \tau_0(\tilde{f}\epsilon_\uparrow)^2\eqno(1.29)
$$
and
$$
m_d\cong \tau_0(\tilde{f}\epsilon_\downarrow)^2\eqno(1.30)
$$
with $\epsilon_\uparrow$ and $\epsilon_\downarrow$ given by (1.13)
and (1.16).

An important feature of the model is to assume that the constant
$\tau_0$ is due to the spontaneous symmetry breaking of a new $T$
odd and $CP$ odd, spin $0$ field $\tau(x)$, which has a vacuum
expectation value given by (1.25). We may assume that the part of
Lagrangian density that contains only $\tau(x)$ is given by
$$
-\frac{1}{2}\bigg(\frac{\partial \tau}{\partial
x_\mu}\bigg)^2-V(\tau)\eqno(1.31)
$$
with
$$
V(\tau)=-\frac{1}{2}\lambda
\tau^2(\tau_0^2-\frac{1}{2}\tau^2)\eqno(1.32)
$$
in which the (renormalized) value of $\lambda$ is positive. This
then yields (1.25). Expanding $V(\tau)$ around its equilibrium value
$\tau=\tau_0$, we have
$$
V(\tau)=-\frac{\lambda}{4}\tau_0^4+\frac{1}{2}m_\tau^2(\tau-\tau_0)^2
+O[(\tau-\tau_0)^3]\eqno(1.33)
$$
with
$$
m_\tau=(2\lambda)^{\frac{1}{2}}\tau_0,\eqno(1.34)
$$
the mass of this new $T$ violating, $C$ violating and $CP$
violating quantum, called timeon. The interaction between
$\tau(x)$ and the quark field can be readily obtained by replacing
the $F=\tau_0f\tilde{f}$ factor of (1.24) with
$$
F=\tau(x)f\tilde{f}.\eqno(1.35)
$$

In this new theory, the well studied mass limit of the standard ($T$
and $CP$ conserving) Higgs boson applies only to the vibrational
modes of the $\gamma_4G_\uparrow$ and $\gamma_4G_\downarrow$ terms
in (1.3). Thus, the timeon mass $m_\tau$ would be different, and
could be much lower.

\newpage

\section*{\Large \sf 2. Two Equivalent Forms of Mass Matrix}

\noindent{\bf 2.1 General Formulation}\\

The "quark" mass matrix ${\cal M}$ and its related Hamiltonian
${\cal H}$ is usually written as
$$
{\cal H}=\Psi^\dag {\cal M} \gamma_4 \Psi\eqno(2.1)
$$
with
$$
{\cal M}={\cal M}^\dag,\eqno(2.2)
$$
denoting a hermitian matrix. Throughout this section, we assume the
Dirac field operator $\Psi$ to have $n$ generation-components, with
$n=3$ for quarks. Decompose $\Psi$ into a sum of left-handed and
right-handed parts:
$$
\Psi={\cal L} + {\cal R}\eqno(2.3)
$$
with
$$
{\cal L}=\frac{1}{2}(1+\gamma_5)\Psi~~~{\sf and}~~~{\cal
R}=\frac{1}{2}(1-\gamma_5)\Psi.\eqno(2.4)
$$
Correspondingly, (2.1) becomes
$$
{\cal H}={\cal L}^\dag{\cal M}\gamma_4{\cal R} +{\cal R}^\dag{\cal
M}\gamma_4{\cal L}.\eqno(2.5)
$$
Assume $n\geq 3$ and ${\cal M}$ to have an imaginary part so that
${\cal H}$ is $T$, $C$ and $CP$ violating.

A different form of an $n$-generation $T$ and $CP$ violating mass
Hamiltonian can be written in the form similar to (1.3), also with a
Dirac operator $\psi$ of $n$ components:
$$
H=\psi^\dag({\cal G}\gamma_4+i{\cal
F}\gamma_4\gamma_5)\psi,\eqno(2.6)
$$
where ${\cal G}$ and ${\cal F}$ are both $n$-dimensional hermitian
matrices,
$$
{\cal G}={\cal G}^\dag~~~{\sf and}~~~{\cal F}={\cal
F}^\dag.\eqno(2.7)
$$
For $n\geq 3$, ${\cal G}$ and ${\cal F}$ both nonzero, the
Hamiltonian $H$ is $T$, $P$ and $CP$ violating. As in (2.3)-(2.4),
we resolve $\psi$ in a similar form:
$$
\psi=L+R\eqno(2.8)
$$
with
$$
L=\frac{1}{2}(1+\gamma_5)\psi~~~{\sf
and}~~~R=\frac{1}{2}(1-\gamma_5)\psi.\eqno(2.9)
$$
Thus, (2.6) becomes
$$
H=L^\dag({\cal G}-i{\cal F})\gamma_4 R+R^\dag({\cal G}+i{\cal
F})\gamma_4L,\eqno(2.10)
$$
different from (2.5).

In the standard model, excluding the mass Hamiltonian, only the
left hand components of $\uparrow$ and $\downarrow$ quarks are
linked by their $W$-interaction. Hence, the right-hand component
${\cal R}$ or $R$ can undergo an independent arbitrary unitary
transformation. Because of this freedom, we can bring (2.10) into
the form (2.5), or vice versa, as we shall see.

To show this, we begin with the form (2.10). Define
$$
M={\cal G}-i{\cal F}\eqno(2.11)
$$
and assume it to be nonsingular (i.e., the eigenvalues of $M^\dag
M$ are all nonzero.) On account of (2.7), the hermitian conjugate
of $M$ is
$$
M^\dag={\cal G}+i{\cal F}.\eqno(2.12)
$$
Since $MM^\dag$ is hermitian, there exists a unitary matrix $V_L$
that can diagonalize $MM^\dag$, with
$$
V_L^\dag MM^\dag V_L=m_D^2=~{\sf Diagonal}.\eqno(2.13)
$$
For every eigenvector $\phi$ of $MM^\dag$ with eigenvalue $\lambda$,
the corresponding vector $M^\dag \phi$ is an eigenvector of $M^\dag
M$ with the same eigenvalue $\lambda$. Thus, $M^\dag M$ can also be
diagonalized by another unitary matrix $V_R$ as
$$
V_R^\dag M^\dag MV_R=m_D^2,\eqno(2.14)
$$
with $m_D^2$ the same diagonal matrix of (2.13).

Multiply (2.13) on the right by $m_D^{-1}$, it follows that
$$
V_L^\dag MV_R=m_D,\eqno(2.15)
$$
provided that we define
$$
V_R=M^\dag V_Lm_D^{-1}.\eqno(2.16)
$$
One can readily see that $V_L$ and $V_R$ thus defined satisfies
$V_L^\dag V_L=1$, $V_R^\dag V_R=1$ as well as (2.13) and (2.14).
Since $R$ can be transformed independently from $L$, we can
transform the $\psi$ field by
$$
L\rightarrow V_LL\eqno(2.17)
$$
and
$$
R\rightarrow V_RR.\eqno(2.18)
$$

Next let us examine the mass matrix ${\cal M}$ of (2.1)-(2.2).
Because ${\cal M}$ is hermitian, it can be diagonalized by a
single unitary transformation $V$, with the left-handed and
right-handed components of the field operator $\Psi$ undergoing
the \underline{same} transformation; i.e., in contract to
(2.17)-(2.18), we have
$$
{\cal L}\rightarrow V{\cal L},\eqno(2.19)
$$
$$
{\cal R}\rightarrow V{\cal R}\eqno(2.20)
$$
and correspondingly
$$
{\cal H}\rightarrow \Psi^\dag m_D\gamma_4\Psi\eqno(2.21)
$$
with $m_D$ being the corresponding diagonal matrix. So far as the
mass matrices are concerned, we regard these two mass Hamiltonians
${\cal H}$ and $H$ as equivalent, if the diagonal matrix $m_D$ of
(2.21) has the same set of eigenvalues as those in (2.15). In this
case, we can without loss of generality set
$$
{\cal M}^2=({\cal G}-i{\cal F})({\cal G}+i{\cal F})\eqno(2.22)
$$
and
$$
V=V_L;\eqno(2.23)
$$
hence (2.13) becomes
$$
V^\dag{\cal M}^2V=m_D^2\eqno(2.24)
$$
and therefore
$$
V^\dag{\cal M} V=m_D.\eqno(2.25)
$$
(Note that ${\cal M}\neq M$ or $M^\dag$, even though ${\cal M}^2=
MM^\dag$.)

We will now discuss theories in which both matrices ${\cal G}$ and
${\cal F}$ are real; i.e.,
$$
{\cal G}={\cal G}^*~~{\sf and}~~{\cal F}={\cal F}^*.\eqno(2.26)
$$
Since ${\cal G}$ and ${\cal F}$ are also hermitian; they must both
be symmetric matrices. In $n$-dimension, each of these matrices
can carry $\frac{1}{2}n(n+1)$ independent real parameters, giving
a total of $n(n+1)$ real parameters. On the other hand, ${\cal M}$
being a single hermitian matrix consists of only $n^2$ real
parameters. Thus, knowing ${\cal G}$ and ${\cal F}$, by using
(2.22), we can always determine uniquely the corresponding ${\cal
M}$, but not the converse, by expanding in power series as
follows.

Decompose the hermitian ${\cal M}$ into its real and imaginary
parts:
$$
{\cal M}=R+iI.\eqno(2.27)
$$
with $R$ and $I$ both real; hence, $R$ is symmetric and $I$
antisymmetric. On account of (2.26), the real part of (2.22) is
$$
R^2-I^2={\cal G}^2+{\cal F}^2,\eqno(2.28)
$$
and the imaginary part is
$$
\{R,~I\}=[{\cal G},~{\cal F}].\eqno(2.29)
$$
In what follows, we assume that ${\cal G}$ and ${\cal F}$ are both
known, as in the case when the mass matrix is given by (1.3). In
addition, ${\cal F}$ can be regarded as small compared to ${\cal
G}$. Hence, we can expend $R$ and $I$ in powers of ${\cal F}$. Write
$$
R={\cal G}+R_2+R_4+R_6+\cdots\eqno(2.30)
$$
and
$$
I=I_1+I_3+I_5+\cdots,\eqno(2.31)
$$
with $R_n$ and $I_m$ to be of the order of ${\cal F}^n$ and ${\cal
F}^m$ respectively. Eqs.(2.28) and (2.29) give
$$
\{{\cal G},~I_1\}=[{\cal G},~{\cal F}],~~~~~~
$$
$$
\{{\cal G},~R_2\}={\cal F}^2+I_1^2,~~~~~
$$
$$
\{{\cal G},~I_3\}=-\{R_2,~I_1\},\eqno(2.32)
$$
$$
\{{\cal G},~R_4\}=\{I_1,~I_3\}-R_2^2,~~{\sf etc.}
$$
As noted before, so far as these mass matrices are concerned, the
two formalisms (2.1) and (2.6) are regarded as equivalent to each
other, provided that (2.28) and (2.29) hold. Then (2.32) gives the
conditions determining the series expansions (2.30)-(2.31) of $R$
and $I$ in terms of ${\cal G}$ and ${\cal F}$.\\

\noindent{\bf 2.2 Application}\\

Next, we apply the above analysis to the special case when
$$
{\cal G}=G_\uparrow~~~{\sf or}~~~G_\downarrow\eqno(2.33)
$$
and $$ {\cal F}=F=F_\uparrow=F_\downarrow\eqno(2.34)
$$
with $G_{\uparrow/\downarrow}$ and $F$ given by (1.4) and (1.24)
respectively. For clarity, we shall suppress the subscript
$\uparrow$ or $\downarrow$ in this section and write (1.13)-(1.18)
as
$$
\epsilon= \left(
\begin{array}{l}
\cos \theta\\
\sin \theta \cos \phi\\
\sin \theta \sin \phi
\end{array}
\right),~~~~~ p = \left(
\begin{array}{l}
-\sin \theta\\
~~\cos \theta \cos \phi\\
~~\cos \theta \sin \phi
\end{array}
\right)
$$
and
$$
P= \left(
\begin{array}{l}
~~~~0\\
-\sin \phi\\
~~\cos \phi
\end{array}
\right). \eqno(2.35)
$$
In order to conform to the sign convention used in (1.13)-(1.18), we
have
$$
\theta=\theta_\uparrow~~~{\sf in~the}~\uparrow~{\sf
sector}\eqno(2.36)
$$
but
$$
\theta=-\theta_\downarrow~~~{\sf in~the}~\downarrow~{\sf
sector}.\eqno(2.37)
$$
Likewise, the unit column matrix $f$ in (1.24) is
$$
f= \left(
\begin{array}{l}
\cos a\\
\sin a \cos b\\
\sin a \sin b
\end{array}
\right).\eqno(2.38)
$$
Define matrix
$$
{\cal G}=\nu\epsilon\tilde{\epsilon}+\mu p\tilde{p} +m
P\tilde{P}\eqno(2.39)
$$
and
$$
{\cal F}=\tau f \tilde{f}\eqno(2.40)
$$
with $\nu,~\mu,~m,~\tau$ all real constants. When $\tau=\tau_0$ and
$\nu=0$, ${\cal F}$ becomes $F$ of (1.24) and ${\cal G}$ can be
either $G_\uparrow$ or $G_\downarrow$ of (1.4). In this section, we
retain the eigenvalue $\nu$ in (2.39) for the formal symmetry of
some of the mathematical expressions, even though $\nu=0$ when we
discuss physical applications of our model in other sections of the
paper.

Let $\vec{{\cal A}}$ be a vector whose $k^{th}$ component is given
by the $(i,~j)^{th}$ component of the commutator between ${\cal G}$
and ${\cal F}$:
$$
[{\cal G},~{\cal F}]_{ij}=\epsilon_{ijk}{\cal A}_k\eqno(2.41)
$$
with $\epsilon _{ijk}=\pm 1$ depending on $(ijk)$ being an even or
odd permutation of $(1,2,3)$, and $0$ otherwise. From (2.35) and
(2.38), we can readily verify that
$$
{\cal A}_k=\tau\bigg(\nu(\hat{\epsilon}\cdot
\hat{f})(\hat{\epsilon}\times\hat{f})+ \mu(\hat{p}\cdot
\hat{f})(\hat{p}\times\hat{f}) +m(\hat{P}\cdot
\hat{f})(\hat{P}\times\hat{f})\bigg)_k.\eqno(2.42)
$$

Next, define a real antisymmetric matrix ${\cal I}$ whose
anti-commutator with ${\cal G}$ is given by
$$
\{{\cal G},~{\cal I}\}=[{\cal G},~{\cal F}].\eqno(2.43)
$$
Let its $(ij)^{th}$ matrix element be written as
$$
{\cal I}_{ij}=\epsilon_{ijk}J_k.\eqno(2.44)
$$
Thus, ${\cal I}$ is the same antisymmetric matrix $I_1$ of (2.31),
i.e.,
$$
{\cal I}=I_1\eqno(2.45)
$$
and (2.43) is the same as the first equation in (2.32).

The corresponding vector $\vec{J}$ is related to $\vec{{\cal A}}$ by
$$
\vec{J}\cdot\hat{\epsilon}=(\mu+m)^{-1}\vec{{\cal A}}\cdot \hat{\epsilon}
$$
$$
\vec{J}\cdot\hat{p}=(m+\nu)^{-1}\vec{{\cal A}}\cdot \hat{p}\eqno(2.46)
$$
and
$$
\vec{J}\cdot\hat{P}=(\nu+\mu)^{-1}\vec{{\cal A}}\cdot \hat{P};
$$
these relations can be readily derived from (2.41)-(2.44) by writing
$G$ as diagonal in the basis $(\hat{\epsilon},~\hat{p},~\hat{P})$.

As in (1.19) and (1.20), we define a real unitary matrix $U_0$
whose columns are $\epsilon,~p$ and $P$ of (2.35); i.e.,
$$
U_0= \left(
\begin{array}{ccc}
\cos \theta & -\sin\theta & 0\\
\sin \theta \cos \phi &\cos \theta \cos \phi & -\sin \phi\\
\sin \theta \sin \phi & \cos \theta \sin \phi & \cos \phi
\end{array}
\right).\eqno(2.47)
$$
The matrix $U_0$ diagonalizes ${\cal G}$, with
$$
{\cal G}' \equiv\tilde{U}_0{\cal G}U_0= \left(
\begin{array}{ccc}
\nu & 0 & 0\\
0 &\mu & 0\\
0 & 0 & m
\end{array}
\right).\eqno(2.48)
$$
It also transforms ${\cal F}$ into
$$
{\cal F}'\equiv \tilde{U}_0{\cal F}U_0 = \tau
f'\tilde{f}'\eqno(2.49)
$$
where
$$
f'=\left(\begin{array}{l}
f_\epsilon\\
f_p\\
f_P
\end{array}
\right)\eqno(2.50)
$$
with
$$
f_\epsilon
=\tilde{\epsilon}f,~~f_p=\tilde{p}f,~~~f_P=\tilde{P}f.\eqno(2.51)
$$
As before,
$$
f_\epsilon^2+f_p^2+f_P^2=1.\eqno(2.52)
$$

By using (2.48)-(2.50), we find that the same $U_0$ also transforms
the matrix ${\cal M}^2$ into
$$
({\cal M}')^2 = \tilde{U}_0{\cal M}^2U_0=({\cal G}')^2+({\cal
F}')^2+i[{\cal G}'~,{\cal F}'],\eqno(2.53)
$$
which is given by
$$
({\cal M}')^2  = \left(
\begin{array}{ccc}
\nu^2+\tau^2 f_\epsilon^2 & \tau[\tau-i(\mu-\nu)]f_\epsilon f_p & \tau[\tau-i(m-\nu)]f_Pf_\epsilon\\
\tau[\tau+i(\mu-\nu)]f_\epsilon f_p&\mu^2+\tau^2f_p^2 & \tau[\tau-i(m-\mu)]f_pf_P\\
\tau[\tau+i(m-\nu)]f_Pf_\epsilon & \tau[\tau+i(m-\mu)]f_pf_P &
m^2+\tau^2f_P^2
\end{array}
\right). \eqno(2.54)
$$
For most of our applications, we are only interested in the case
$\nu=0$. Define
$$
{\cal M}_0' \equiv \lim_{\nu=0}{\cal M}'\eqno(2.55)
$$
and let $\lambda_1^2,~\lambda_2^2,~\lambda_3^2$ be the eigenvalues
of $({\cal M}_0')^2$. From (2.54) and (2.55) we have
$$
\lambda_1^2+\lambda_2^2+\lambda_3^2=m^2+\mu^2+\tau^2\eqno(2.56)
$$
and
$$
\lambda_1^2\lambda_2^2\lambda_3^2=|({\cal
M}'_0)^2|=\tau^2f_\epsilon^4\mu^2m^2,\eqno(2.57)
$$
These eigenvalues are also the solution $\lambda^2$ of the cubic
equation
$$
|({\cal M}'_0)^2-\lambda^2|=|({\cal
M}'_0)^2|+A\lambda^2+B\lambda^4-\lambda^6=0\eqno(2.58)
$$
where
$$
A=-\mu^2m^2-\tau^2[m^2(1-f_P^2)^2+\mu^2(1-f_p^2)^2+2m\mu
f_p^2f_P^2]\eqno(2.59)
$$
and
$$
B=m^2+\mu^2+\tau^2.\eqno(2.60)
$$
In the limit $\tau\rightarrow 0$, we see from (2.57)-(2.60) that
the two heavier masses become $\mu$ and $m$, while the lightest
mass is proportional to $\tau$. We shall explore this limit
further in the next section.

\newpage

\section*{\Large \sf 3. Perturbative Solution and Jarlskog Invariant}

\noindent{\bf 3.1 Perturbation Series}\\

In this section we return to the mass Hamiltonian (1.3) and
calculate its eigenstates by using $G\gamma_4$ as the zeroth order
Hamiltonian and $iF\gamma_4\gamma_5$ as the perturbation. From the
discussions given in the last section, we see that this is
identical to the problem of finding the eigenstates of $({\cal
M}_0')^2$ regarding $\tau$ as the small parameter. Using
(2.54)-(2.55), we may write
$$
({\cal M}_0')^2 = \left(
\begin{array}{ccc}
0 & 0 & 0\\
0 &\mu^2 & 0\\
0 & 0 & m^2
\end{array}
\right)+h+O(\tau^2)\eqno(3.1)
$$
with
$$
h = \tau \left(
\begin{array}{ccc}
0 & -i\mu f_\epsilon f_p & -im f_P f_\epsilon\\
i\mu f_\epsilon f_p &0 & -i(m-\mu)f_p f_P\\
im f_P f_\epsilon & i(m-\mu)f_p f_P & 0
\end{array}
\right).\eqno(3.2)
$$
To first order in $h$, the eigenstates of $({\cal M}_0')^2$ can be
readily obtained. For applications to physical quarks, we need
only to identify that $f_\epsilon$, $f_p$ and $f_P$ are replaced
by
$$
(f_\epsilon)_{\uparrow/\downarrow}=\tilde{f}\epsilon_{\uparrow/\downarrow},
$$
$$
(f_p)_{\uparrow/\downarrow}=\tilde{f}p_{\uparrow/\downarrow}\eqno(3.3)
$$
and
$$
(f_P)_{\uparrow/\downarrow}=\tilde{f}P_{\uparrow/\downarrow}.
$$
Likewise, neglecting $O(\tau^2)$ corrections, we can relate
$\mu^2$ and $m^2$ to (quark~mass)$^2$ by
$$
\mu^2_\uparrow = m_c^2,~~{\sf ~~~}~~m_\uparrow^2=m_t^2
$$
$$
\mu^2_\downarrow = m_s^2,~~{\sf
~~~}~~m_\downarrow^2=m_b^2\eqno(3.4)
$$
and set
$$
\tau=\tau_0.\eqno(3.5)
$$
Thus, to $O(\tau_0)$, in the $\uparrow$ sector the state vectors
of $u,~c,~t$ are related to those of
$\epsilon_\uparrow,~p_\uparrow$ and $P_\uparrow$ by
$$
\left(\begin{array}{l}
u\\
c\\
t
\end{array}
\right)=\left(
\begin{array}{ccc}
(\epsilon_\uparrow|u)& (p_\uparrow|u) & (P_\uparrow|u)\\
(\epsilon_\uparrow|c) &(p_\uparrow|c)&(P_\uparrow|c)\\
(\epsilon_\uparrow|t) &(p_\uparrow|t)&(P_\uparrow|t)
\end{array}\right)\left(\begin{array}{l}
\epsilon_\uparrow\\
p_\uparrow\\
P_\uparrow
\end{array}
\right)\eqno(3.6)
$$
with
$$
(\epsilon_\uparrow|u)=1+O(\tau_0^2),~~~~~~~~~~~~~~\eqno(3.7)
$$
$$
(\epsilon_\uparrow|c)=-i\frac{\tau_0}{m_c}(f_\epsilon
f_p)_\uparrow+O(\tau_0^2),\eqno(3.8)
$$
$$
(\epsilon_\uparrow|t)=-i\frac{\tau_0}{m_t}(f_\epsilon
f_P)_\uparrow+O(\tau_0^2),\eqno(3.9)
$$

$$
(p_\uparrow|u)=-i\frac{\tau_0}{m_c}(f_\epsilon
f_p)_\uparrow+O(\tau_0^2),\eqno(3.10)
$$
$$
(p_\uparrow|c)=1+O(\tau_0^2),~~~~~~~~~~~~~~\eqno(3.11)
$$
$$
~~~~~~~~(p_\uparrow|t)=-i\frac{\tau_0}{m_t+m_c}(f_p
f_P)_\uparrow+O(\tau_0^2),\eqno(3.12)
$$

$$
~(P_\uparrow|u)=-i\frac{\tau_0}{m_t}(f_\epsilon
f_P)_\uparrow+O(\tau_0^2),\eqno(3.13)
$$
$$
~~~~~~~~(P_\uparrow|c)=-i\frac{\tau_0}{m_t+m_c}(f_p
f_P)_\uparrow+O(\tau_0^2)\eqno(3.14)
$$
and
$$
(P_\uparrow|t)=1+O(\tau_0^2).~~~~~~~~~~~~~\eqno(3.15)
$$
Likewise, for the $\downarrow$ sector, we may write
$$
\left(\begin{array}{l}
d\\
s\\
b
\end{array}
\right)=\left(
\begin{array}{ccc}
(\epsilon_\downarrow|d)& (p_\downarrow|d) &(P_\downarrow|d)\\
(\epsilon_\downarrow|s) &(p_\downarrow|s)&(P_\downarrow|s)\\
(\epsilon_\downarrow|b) &(p_\downarrow|b)&(P_\downarrow|b)
\end{array}\right)\left(\begin{array}{l}
\epsilon_\downarrow\\
p_\downarrow\\
P_\downarrow
\end{array}
\right)\eqno(3.16)
$$
with
$$
(\epsilon_\downarrow|d)=1+O(\tau_0^2),~~~~~~~~~~~~~~\eqno(3.17)
$$
$$
(\epsilon_\downarrow|s)=-i\frac{\tau_0}{m_s}(f_\epsilon
f_p)_\downarrow+O(\tau_0^2),\eqno(3.18)
$$
$$
(\epsilon_\downarrow|b)=-i\frac{\tau_0}{m_b}(f_\epsilon
f_P)_\downarrow+O(\tau_0^2),\eqno(3.19)
$$

$$
(p_\downarrow|d)=-i\frac{\tau_0}{m_s}(f_\epsilon
f_p)_\downarrow+O(\tau_0^2),\eqno(3.20)
$$
$$
(p_\downarrow|s)=1+O(\tau_0^2),~~~~~~~~~~~~~~\eqno(3.21)
$$
$$
~~~~~~~~(p_\downarrow|b)=-i\frac{\tau_0}{m_b+m_s}(f_p
f_P)_\downarrow+O(\tau_0^2),\eqno(3.22)
$$

$$
~(P_\downarrow|d)=-i\frac{\tau_0}{m_b}(f_\epsilon
f_P)_\downarrow+O(\tau_0^2),\eqno(3.23)
$$
$$
~~~~~~~~(P_\downarrow|s)=-i\frac{\tau_0}{m_b+m_s}(f_p
f_P)_\downarrow+O(\tau_0^2)\eqno(3.24)
$$
and
$$
(P_\downarrow|b)=1+O(\tau_0^2).~~~~~~~~~~~~~\eqno(3.25)
$$

\noindent{\bf 3.2 Jarlskog Invariant}\\

Write the CKM matrix as
$$
U_{CKM} = \left(
\begin{array}{ccc}
U_{11}& U_{12} &U_{13}\\
U_{21}& U_{22} &U_{23}\\
U_{31}& U_{32} &U_{33}
\end{array}\right).\eqno(3.26)
$$
Following Jarlskog[4], we introduce
$$
S_1=U_{11}^*U_{12},~~~S_2=U_{21}^*U_{22},~~~S_3=U_{31}^*U_{32}\eqno(3.27)
$$
and define
$$
{\cal J}=ImS_1^*S_2.\eqno(3.28)
$$
By using (3.27) we see that
$$
{\cal
J}=Im\bigg[(U_{11}U_{22})(U_{12}^*U_{21}^*)\bigg],\eqno(3.29)
$$
Because of unitarity of the CKM matrix,
$$
S_1+S_2+S_3=0.\eqno(3.30)
$$
Therefore, ${\cal J}$ is equal to twice the area of the triangle
whose sides are $S_1,~S_2$ and $S_3$. Furthermore, from the
explicit form of ${\cal J}$ given by (3.29), we see that ${\cal
J}$ is symmetric with respect to the interchange between the row
and column indices of the CKM matrix. It follows then in deriving
${\cal J}$, we may use the elements of either any two columns or
any two rows of the CKM matrix.

It is convenient to denote $(U_{CKM})_0$ of (1.21) simply as $V$,
with
$$
V\equiv (U_{CKM})_0 = \left(
\begin{array}{ccc}
V_{11}& V_{12} &V_{13}\\
V_{21}& V_{22} &V_{23}\\
V_{31}& V_{32} &V_{33}
\end{array}\right).\eqno(3.31)
$$
In terms of the state vectors
$\epsilon_\uparrow,~p_\uparrow.~P_\uparrow$ and
$\epsilon_\downarrow,~p_\downarrow,~P_\downarrow$ of
(1.13)-(1.18), we can also write $V$ as
$$
V=\left(
\begin{array}{ccc}
(\epsilon_\uparrow|\epsilon_\downarrow)& (\epsilon_\uparrow|p_\downarrow) & (\epsilon_\uparrow|P_\downarrow)\\
(p_\uparrow|\epsilon_\downarrow) &(p_\uparrow|p_\downarrow)&(p_\uparrow|P_\downarrow)\\
(P_\uparrow|\epsilon_\downarrow)
&(P_\uparrow|p_\downarrow)&(P_\uparrow|P_\downarrow)
\end{array}\right).\eqno(3.32)
$$
Likewise, the CKM matrix is given by
$$
U_{CKM}= \left(
\begin{array}{ccc}
(u|d)& (u|s) & (u|b)\\
(c|d) &(c|s)&(c|b)\\
(t|d) &(t|s)&(t|b)
\end{array}\right)
$$
$$
~~~~=V+i\tau_0W+O(\tau_0^2)\eqno(3.33)
$$
where the matrix elements of $W$ are derived from (3.7)-(3.15) and
(3.17)-(3.25). Using the perturbative solution of Sec. 3.1, we can
readily express the matrix elements of $U_{CKM}$ in terms of the
corresponding ones of $V$. The Jarlskog invariant can then be
evaluated by using (3.29).

As will be shown in the Appendix, the result, accurate to the
first power of $\tau_0$, is
$$
{\cal J}=\tau_0\bigg[\frac{(f_\epsilon f_p)_\downarrow}{m_s}A_s+
\frac{(f_\epsilon f_P)_\downarrow}{m_b}A_b +\frac{(f_p
f_P)_\downarrow}{m_s+m_b}B_\downarrow
$$
$$
+ \frac{(f_\epsilon f_p)_\uparrow}{m_c}A_c+\frac{(f_\epsilon
f_P)_\uparrow}{m_t}A_t+ \frac{(f_p
f_P)_\uparrow}{m_c+m_t}B_\uparrow\bigg]\eqno(3.34)
$$
where
$$
A_s=-V_{13}V_{23}V_{33}\cong -2\cdot 10^{-4},\eqno(3.35)
$$
$$
A_b=-V_{12}V_{22}V_{32}\cong 8.8\cdot 10^{-3},~\eqno(3.36)
$$
$$
B_\downarrow=-V_{11}V_{21}V_{31}\cong 1.10\cdot
10^{-3},\eqno(3.37)
$$
$$
A_c=V_{31}V_{32}V_{33}\cong -2\cdot 10^{-4},~~\eqno(3.38)
$$
$$
A_t=V_{21}V_{22}V_{23}\cong -8.8\cdot 10^{-3}\eqno(3.39)
$$
and
$$
B_\uparrow=V_{11}V_{12}V_{13}\cong 1.10\cdot 10^{-3}.\eqno(3.40)
$$

From the definition (3.29) and (3.34), these coefficients
$A_s,~\cdots,~B_\uparrow$ are all products of four factors of
$V_{ij}$. As will be shown in the Appendix, because $V$ is a real
orthogonal matrix, these quartic products can all be reduced to
triple products given by (3.35)-(3.40). Since $m_c>>m_s$ and
$m_t>>m_b$, we can, as an approximation, neglect the terms related
to the up sector in (3.34).

\newpage

\section*{\Large \sf 4. Determination of $\tau_0$ and $f$}

\noindent{\bf 4.1 A Special coordinate system}\\

For the $\uparrow$ quarks, the parameters $\lambda_1,~\lambda_2$
and $\lambda_3$ in (2.56)-(2.60) are related to the quark masses
by
$$
\lambda_1=m_u,~~\lambda_2=m_c~~{\sf and}~~\lambda_3=m_t.\eqno(4.1)
$$
Likewise, $f_\epsilon$ is
$$
(f_\epsilon)_\uparrow=\tilde{f}\epsilon_\uparrow\eqno(4.2)
$$
with $\epsilon_\uparrow$ given by (1.13) and $f$ the unit
directional vector of (1.24). We work to leading order in
$\tau_0$. From (2.57), setting $\lambda_2=\mu$ and $\lambda_3=m$
we have
$$
m_u=\tau_0(\tilde{f}\epsilon_\uparrow)^2.\eqno(4.3)
$$
Likewise, for the $\downarrow$ quarks
$$
m_d=\tau_0(\tilde{f}\epsilon_\downarrow)^2.\eqno(4.4)
$$
It is convenient to introduce a special coordinate system in which
$$
\epsilon_\downarrow=\left(\begin{array}{l}
1\\
0\\
0
\end{array}
\right)~~{\sf and}~~ \epsilon_\uparrow=\left(\begin{array}{l}
~~\cos\theta_c\\
-\sin\theta_c\\
~~~~0
\end{array}
\right).\eqno(4.5)
$$
Since $p_\downarrow$ and $P_\downarrow$ are both $\bot
\epsilon_\downarrow$, we may write
$$
p_\downarrow=\left(\begin{array}{l}
~~~~0\\
-\cos\gamma\\
~~\sin\gamma
\end{array}
\right)~~{\sf and}~~ P_\downarrow=\left(\begin{array}{l}
~~~~0\\
-\sin\gamma\\
-\cos\gamma
\end{array}
\right).\eqno(4.6)
$$
Furthermore, we shall set the zeroth order CKM matrix
$(U_{CKM})_0$ of (1.21) to be
$$
(U_{CKM})_0=\left(
\begin{array}{ccc}
(\epsilon_\uparrow|\epsilon_\downarrow)& (\epsilon_\uparrow|p_\downarrow) & (\epsilon_\uparrow|P_\downarrow)\\
(p_\uparrow|\epsilon_\downarrow) &(p_\uparrow|p_\downarrow)&(p_\uparrow|P_\downarrow)\\
(P_\uparrow|\epsilon_\downarrow)
&(P_\uparrow|p_\downarrow)&(P_\uparrow|P_\downarrow)
\end{array}\right)
$$
$$
=\left(
\begin{array}{ccc}
~~.974& ~~~~.227 & ~~5\cdot 10^{-3}\\
-.227& ~~~~.973 & ~.04\\
5\cdot 10^{-3}& -.04 & ~~~.999
\end{array}\right)+O(1\cdot 10^{-3}).\eqno(4.7)
$$
Thus, with the same accuracy of $O(10^{-3})$,
$$
\cos \theta_c=.974~~{\sf and}~~\sin\theta_c=.227.\eqno(4.8)
$$
Likewise, from $(\epsilon_\uparrow|P_\downarrow)=5\cdot 10^{-3}$ in
(4.7), in accordance with (4.5), (4.6) and
$$
(\epsilon_\uparrow|P_\downarrow)=\sin\theta_c\sin\gamma,\eqno(4.9)
$$
we find
$$
\sin\gamma=2.2\cdot 10^{-2},\eqno(4.10)
$$
which together with (4.5) and (4.6) give the coordinate system
defined by $(\epsilon_\downarrow,~p_\downarrow,P_\downarrow)$.
Eq.(4.7) then, in turn, determines the corresponding coordinate
system  $(\epsilon_\uparrow,~p_\uparrow,P_\uparrow)$.

Next, we shall determine the parameters $\tau_0$ and the directional
angles $\alpha$ and $\beta$ of the unit vector
$$
f= \left(
\begin{array}{l}
\sin\alpha\cos\beta\\
\sin\alpha\sin\beta\\
\cos\alpha
\end{array}
\right)\eqno(4.11)
$$
in the coordinate system defined by (4.5)-(4.6).\\

\noindent{\bf 4.2 Determination of $\beta$}\\

From (4.5) and (4.11), we have
$$
\tilde{f}\epsilon_\downarrow=\sin\alpha\cos\beta\eqno(4.12)
$$
and
$$
\tilde{f}\epsilon_\uparrow=\sin\alpha\cos(\beta+\theta_c).\eqno(4.13)
$$
Thus, on account of (4.3) and (4.4),
$$
\frac{\cos^2(\beta+\theta_c)}{\cos^2\beta}=\frac{m_u}{m_d}\eqno(4.14)
$$
and therefore
$$
\frac{\cos(\beta+\theta_c)}{\cos\beta}=\pm\bigg(\frac{m_u}{m_d}\bigg)^{\frac{1}{2}}.\eqno(4.15)
$$
assuming
$$
\frac{m_u}{m_d}\cong \frac{1}{2},\eqno(4.16)
$$
we find two solutions for $\beta$:
$$
\beta\cong 48^0~50'\eqno(4.17)
$$
or
$$
\beta\cong 82^0~20'.\eqno(4.18)
$$

\noindent{\bf 4.3 Determination of $\alpha$ and $\tau_0$}\\

We shall first determine the parameter $\alpha$ by using the
Jarlskog invariant
$$
{\cal J}=3.08\cdot 10^{-5}.\eqno(4.19)
$$
Define
$$
F=10^2{\cal J}m_b/\tau_0.\eqno(4.20)
$$
From (4.4), (4.5) and (4.11), we have
$$
m_d=\tau_0\sin^2\alpha\cos^2\beta\eqno(4.21)
$$
and therefore
$$
F=10^2{\cal J}(m_b/m_d)\sin^2\alpha\cos^2\beta.\eqno(4.22)
$$
For definiteness, we shall set the various quark masses as
$$
m_d \cong 5MeV,~~~~~~~m_u\cong 2.5MeV
$$
$$
m_s\cong 95MeV,~~~~~~~m_c\cong 1.25GeV
$$
$$
m_b\cong 4.2GeV~~{\sf and}~~m_t\cong 175GeV,\eqno(4.23)
$$
consistent with the Particle Data Group values[5]. Thus, (4.22)
becomes
$$
F(\alpha,~\beta)\cong 2.6\sin^2\alpha\cos^2\beta,\eqno(4.24)
$$

On the otherhand, from (3.34) and by using the numerical values for
$A_s,~A_b,~\cdots,~B_\uparrow$ of (3.35)-(3.40) together with the
various quark masses given above, the same function
$F(\alpha,~\beta)$ is also
$$
F(\alpha,~\beta)\cong -0.88(f_\epsilon
f_p)_\downarrow~~-0.067~(f_\epsilon f_p)_\uparrow
$$
$$
~~~~~~~~~~~~~ +0.88(f_\epsilon f_P)_\downarrow~~-0.021~(f_\epsilon
f_P)_\uparrow
$$
$$
~~~~~~~~~~~~~~ +0.11(f_p f_P)_\downarrow~~+0.0026(f_p
f_P)_\uparrow.\eqno(4.25)
$$
As an approximation, we may neglect the contributions of the
$\uparrow$ sector. Combining (4.24) with (4.25), we find
$$
2.6 \sin^2\alpha\cos^2\beta\cong 0.88
f_{\epsilon_\downarrow}[f_{P_\downarrow}-f_{p_\downarrow}]
+0.11(f_pf_P)_\downarrow\eqno(4.26)
$$
with
$$
f_{\epsilon_\downarrow}=\tilde{f}\epsilon_\downarrow\eqno(4.27)
$$
given by (4.12),
$$
f_{p_\downarrow}=\tilde{f}p_\downarrow=-\sin\alpha\sin\beta\cos\gamma+\cos\alpha\sin\gamma\eqno(4.28)
$$
and
$$
f_{P_\downarrow}=\tilde{f}P_\downarrow=-\sin\alpha\sin\beta\sin\gamma-\cos\alpha\cos\gamma.\eqno(4.29)
$$
By using $\gamma$ from (4.10),
$$
\beta\cong48^0~50'
$$
from (4.17), we find
$$
\alpha\cong -36^0~10'.\eqno(4.30)
$$
From (4.4) and (4.12), we have
$$
\tau_0=m_d/(\sin^2\alpha\cos^2\beta).\eqno(4.31)
$$
The above values of $\alpha,~\beta$ and $m_d\cong 5MeV$ give
$$
\tau_0\cong 33MeV.\eqno(4.32)
$$

On the otherhand, the alternative solution $\beta\cong 82^0~20'$ of
(4.18) leads to a much larger value $\tau_0\sim 5.5GeV$. Such a
large value invalidates the small $\tau_0$ approximation. Thus, we
focus only on the solution (4.32) in this paper.

\section*{\Large \sf 5. Remarks}

While the discovery[6] of $T$ and $CP$ violation was made 44 years
ago, at present very little is known about its origin. The totality
of our knowledge can be summarized by a single dimensionless small
number, the Jarlskog invariant of the CKM matrix. In this paper, we
present a simple dynamic model. The vibration of the spin $0$ field
$\tau(x)$ around its equilibrium value $\tau_0$ gives a new quantum,
the timeon. Since $\tau(x)$ is $T$ odd and $CP$ odd, viewed from the
context of the Standard Model[1], the timeon field violates
maximally its $T$ conservation and $CP$  conservation. As mentioned
in the Introduction, the timeon mass $m_\tau$ would be different and
might be much lower than the mass limit of the ($T$ and $CP$
conserving) Higgs boson in the standard model.

In order to detect timeon, a useful signal might be the observation
in any reaction, such as
$$
p+p\rightarrow N+N'+\tau+\cdots,\eqno(5.1)
$$
"$T$ violating" kinematic variables, like
$$
(\vec{k}_N\times \vec{k}_{N'})\cdot \vec{k}\tau.\eqno(5.2)
$$
Because of the final state interaction, a mere observation of such
variables would not be decisive; its amplitude must be above a
certain kinematic limit[7]. Nevertheless these could be useful
signals.

\newpage

\section*{\Large \sf Appendix}

\noindent{\bf A.1 Definitions, Corollaries and Conventions}\\

Let $V=(V_{i\alpha})$ be a $3\times 3$ real orthogonal matrix with
positive determinant, and indices $i$ and $\alpha=1,~2$ and $3$;
hence,
$$
V=V^*,~~V^{-1}=\tilde{V}~~{\sf and}~~|V|=1.\eqno(A.1)
$$
Given any value of $i$, define $i',~i''$ by
$$
\epsilon_{ii'i''}=1,\eqno(A.2)
$$
so that $(i,~i',~i'')$ is a cyclic permutation of $(1,~2,~3)$. We
shall use the same definition for each of such similar indices
$j,~k,~l,~\alpha,~\beta,~\gamma$. Thus, for a given $j$, the
corresponding $j'$ and $j''$ satisfy
$$
\epsilon_{jj'j''}=1,\eqno(A.3)
$$
and likewise
$$
\epsilon_{kk'k''}=\epsilon_{ll'l''}=\epsilon_{\alpha\alpha'\alpha''}
=\epsilon_{\beta\beta'\beta''}=\epsilon_{\gamma\gamma'\gamma''}=1.
$$
Furthermore, for any pair $i,~\alpha$, we have by the expression for
$V^{-1}$
$$
\left|
\begin{array}{cc}
V_{i'\alpha'} & V_{i'\alpha''}\\
V_{i''\alpha'} & V_{i''\alpha''}
\end{array}
\right| =|V|(V^{-1})_{\alpha i}=V_{i\alpha}\eqno(A.4)
$$
on account of (A.1). This identity will enable us to reduce certain
quartic products of $V_{i\alpha}$ to triple products of
$V_{i\alpha}$, as we shall see.\\

\noindent{\bf A.2 Jarlskog Invariant}\\

Similar to the relation between $V$ of (3.32) and
$$
U=U_{CKM}\eqno(A.5)
$$
of (3.33), we define $W=(W_{i\alpha})$ through
$$
U=V+i\tau_0W\eqno(A.6)
$$
where $W$ has the form
$$
W_{i\alpha}=\sum_{j \neq i}F_{ij}V_{j\alpha} -\sum_{\beta
\neq\alpha}{\cal F}_{\alpha\beta}V_{i\beta}.\eqno(A.7)
$$
For our purpose here, it is necessary to specify only that $F$ and
${\cal F}$ are real and symmetric; i.e.,
$$
F_{ij}=F_{ji}=F_{ij}^*\eqno(A.8)
$$
and
$$
{\cal F}_{\alpha\beta}={\cal F}_{\beta\alpha}={\cal
F}_{\alpha\beta}^*.\eqno(A.9)
$$
We also define for any particular pair of indices $k$ and $\gamma$,
$$
J= U_{k\gamma} U_{k'\gamma'}U_{k\gamma'}^* U_{k'\gamma}^*\eqno(A.10)
$$
and
$$
J_0= V_{k\gamma} V_{k'\gamma'}V_{k\gamma'}^*
V_{k'\gamma}^*.\eqno(A.11)
$$
Thus, substituting (A.6) into (A.10), we find, to first order in
$\tau_0$,
$$
J-J_0=i\tau_0\Delta_{k\gamma}\eqno(A.12)
$$
where
$$
\Delta_{k\gamma}=J_0\bigg(\frac{W_{k\gamma}}{V_{k\gamma}}+
\frac{W_{k'\gamma'}}{V_{k'\gamma'}}-\frac{W_{k\gamma'}}{V_{k\gamma'}}
-\frac{W_{k'\gamma}}{V_{k'\gamma}}\bigg).\eqno(A.13)
$$
Note that $k,~\gamma$  are subject to the cyclic convention typified
by (A.2). [It will turn out that $\Delta_{k\gamma}$ is independent
of the choice of $k$ and $\gamma$, even though this is not assumed.]

By substituting (A.7) into (A.13), we must obtain an expression of
the form
$$
\Delta_{k\gamma}=\sum_l A_{l''}F_{ll'}+ \sum_\lambda {\cal
A}_{\lambda''}{\cal F}_{\lambda\lambda'}\eqno(A.14)
$$
where the $A's$ and ${\cal A}'s$ are to be determined. From (A.7),
(A.13) and (A.14), we see that each $A$ is made of terms having
the form $J_0V_{j\alpha}/V_{i\alpha}$. Consider $A_k$: in (A.14)
we must put $l''=k$; hence $l,~l'$ are $k',~k''$ in some order.
Thus in (A.7), $i$ is either $k'$ or $k''$. But the index $k''$
does not occur in (A.13). Therefore, we have $i=k', j=k''$ and
$\alpha=\gamma$ or $\gamma'$. Therefore,
$$
A_k=J_0\bigg(0+\frac{V_{k''\gamma'}}{V_{k'\gamma'}}-0-
\frac{V_{k''\gamma}}{V_{k'\gamma}}\bigg)
$$
$$
=V_{k\gamma}V_{k\gamma'}\left|
\begin{array}{cc}
V_{k'\gamma} & V_{k'\gamma'}\\
V_{k''\gamma} & V_{k''\gamma'}
\end{array}
\right| =V_{k\gamma}V_{k\gamma'}V_{k\gamma''}\eqno(A.15)
$$
on account of (A.4). Likewise,
$$
A_{k'}=V_{k'\gamma}V_{k'\gamma'}V_{k'\gamma''}.\eqno(A.16)
$$

For $k''$, the calculation is different, even though the result will
be similar. Note that in (A.14) $l,~l'$ can be $k,~k'$ in either
order, so that in (A.7) $i$ and $j$ can also be $k,~k'$ in either
order. Thus, we now have four terms instead of two:
$$
A_{k''}=J_0\bigg(\frac{V_{k'\gamma}}{V_{k\gamma}}+\frac{V_{k\gamma'}}{V_{k'\gamma'}}
-\frac{V_{k'\gamma'}}{V_{k\gamma'}}-
\frac{V_{k\gamma}}{V_{k'\gamma}}\bigg)
$$
$$
=V_{k'\gamma}V_{k'\gamma'}\left|
\begin{array}{cc}
V_{k'\gamma} & V_{k'\gamma'}\\
V_{k\gamma} & V_{k\gamma'}
\end{array}
\right| + V_{k\gamma}V_{k\gamma'}\left|
\begin{array}{cc}
V_{k\gamma'} & V_{k\gamma}\\
V_{k'\gamma'} & V_{k'\gamma}
\end{array}
\right|
$$
$$
=-V_{k'\gamma}V_{k'\gamma'}V_{k''\gamma''}-V_{k\gamma}V_{k\gamma'}V_{k\gamma''}
=+V_{k''\gamma}V_{k''\gamma'}V_{k''\gamma''}\eqno(A.17)
$$
since $V$ is orthogonal.

Similar formulas are obtained for ${\cal A}_\gamma$, ${\cal
A}_\gamma'$, ${\cal A}_\gamma''$ with a change of sign. Since
(A.15)-(A.17) all have the same form, the choice of $k,~\gamma$ in
(A.10) and (A.11) is immaterial  and we have
$$
J=J_0+i\tau_0\Delta\eqno(A.18)
$$
where
$$
\Delta=\sum_l F_{ll'}\prod_\alpha V_{l''\alpha}-\sum_\lambda {\cal
F}_{\lambda\lambda'}\prod_i V_{i\lambda''}.\eqno(A.19)
$$

\noindent{\bf A.3 Applications to quarks}\\

By consulting (3.7)-(3.15) and (3.17)-(3.25), we find that
$$
F_{ij}=\frac{f_{i}f_{j}}{m_i+m_j},~~~ {\cal F} _{\alpha\beta}=\frac{
\bar{f}_{\alpha}\bar{f}_\beta}{\mu_\alpha+\mu_\beta}\eqno(A.20)
$$
where
$$
f_1=(f_\epsilon)_\uparrow,~~f_2=(f_p)_\uparrow,~~f_3=(f_P)_\uparrow,
$$
$$
\bar{f}_1=(f_\epsilon)_\downarrow,~~
\bar{f}_2=(f_p)_\downarrow,~~\bar{f}_3=(f_P)_\downarrow\eqno(A.21)
$$
with
$$
m_1=m_u=0,~~m_2=m_c,~~m_3=m_t,
$$
$$
\mu_1=m_d=0,~~\mu_2=m_s,~~\mu_3=m_b.\eqno(A.22)
$$
Then (3.34) is seen to be (A.14) if we identify
$$
A_c=A_3,~~A_t=A_2,~~B_\uparrow=A_1,
$$
$$
A_s={\cal A}_3,~~A_b={\cal A}_2~~{\sf and}~~B_\downarrow={\cal
A}_1.\eqno(A.23)
$$
With these identifications, (3.38)-(3.40) are (A.15)-(A.17) and
(3.35)-(3.37) are the corresponding formulas for ${\cal A}_1,~{\cal
A}_2$ and ${\cal A}_3$.

\vspace{2cm}

\centerline{\epsfig{file=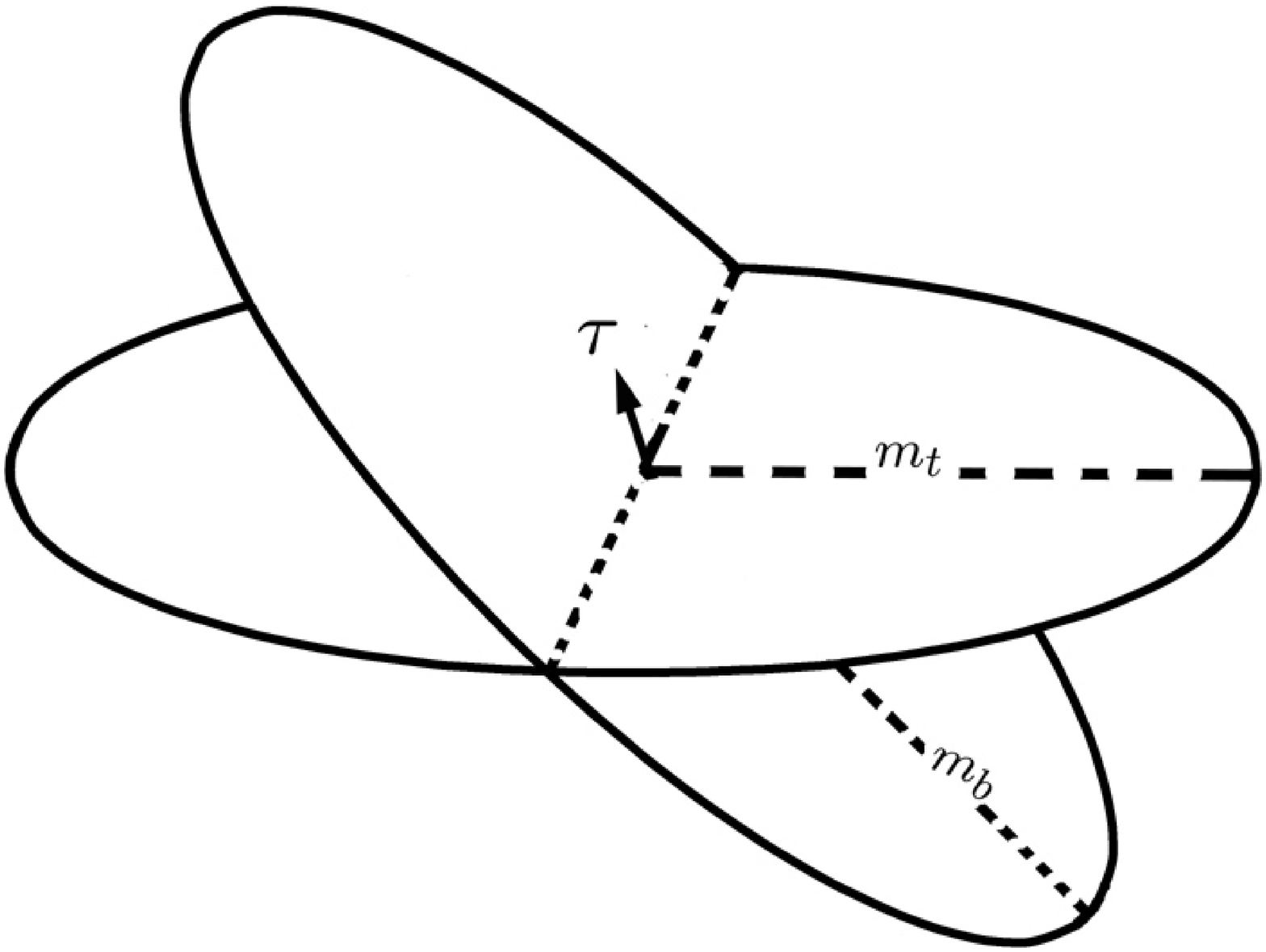,width=10cm}}

\vspace{.5cm}

\noindent Figure~1. A schematic drawing of the mass matrix
$M_{\uparrow/\downarrow}=G_{\uparrow/\downarrow}\gamma_4 +
iF\gamma_4 \gamma_5$. The vibration of $\tau(x)$ is timeon

\newpage

\section*{\Large \sf References}

\noindent [1] S. Weinberg, Phys. Rev. Lett. {\bf 19}(1967)1264\\

\noindent [2] R. Friedberg and T. D. Lee, Ann. Phys. {\bf 323}(2008)1677\\

\noindent [3] T. D. Lee, Phys. Reports {\bf 9}(1974)143\\

\noindent [4] C. Jarlskog, Phys. Rev. {\bf D35}(1987)1685\\

\noindent [5] S. Eidelman et.al. Particle Data Group, Phys. Lett. {\bf B592}(2004)1\\

\noindent [6] J. H. Christenson, J. W. Cronin, V. L. Fitch and R.
Turlay,

~~~~Phys. Rev. Lett. {\bf 13}(1964)138\\

\noindent [7] T. D. Lee and C. N. Yang, Elementary Particles and
Weak Interactions,

~~~~Brookhaven Laboratory Report {\bf 443-T-91}(1957);

~~~~p271 of Vol.~2, T. D. Lee: Selected Papers, Birkhauser Boston Inc.\\
\hspace*{2cm}(1986).

\end{document}